\documentclass[english,aps,prl,twocolumn,superscriptaddress,address,showacknowledgments,longbibliography]{revtex4-1}
\usepackage[T1]{fontenc}
\usepackage[latin9]{inputenc}
\usepackage{babel}
\usepackage{amssymb}
\usepackage{graphicx}
\usepackage{esint}
\usepackage[unicode=true,pdfusetitle,
 bookmarks=true,bookmarksnumbered=false,bookmarksopen=false,
 breaklinks=false,pdfborder={0 0 1},backref=false,colorlinks=false]
 {hyperref}

\makeatletter
\@ifundefined{textcolor}{}
{%
 \definecolor{BLACK}{gray}{0}
 \definecolor{WHITE}{gray}{1}
 \definecolor{RED}{rgb}{1,0,0}
 \definecolor{GREEN}{rgb}{0,1,0}
 \definecolor{BLUE}{rgb}{0,0,1}
 \definecolor{CYAN}{cmyk}{1,0,0,0}
 \definecolor{MAGENTA}{cmyk}{0,1,0,0}
 \definecolor{YELLOW}{cmyk}{0,0,1,0}
}

\makeatother

\begin{document}

\title{Intracavity squeezing can enhance quantum-limited optomechanical
position detection through de-amplification}

\author{V. Peano}

\affiliation{Institute for Theoretical Physics, Friedrich-Alexander-Universität
Erlangen-Nürnberg, Staudtstr. 7, 91058 Erlangen, Germany}

\author{H. G. L. Schwefel}

\affiliation{Institute of Optics, Information and Photonics, Friedrich-Alexander-Universität
Erlangen-Nürnberg, Staudtstr. 7, 91058 Erlangen, Germany}

\affiliation{Max Planck Institute for the Science of Light, Günther-Scharowsky-Straße
1 Bau 24, 91058 Erlangen, Germany}

\author{Ch. Marquardt}

\affiliation{Institute of Optics, Information and Photonics, Friedrich-Alexander-Universität
Erlangen-Nürnberg, Staudtstr. 7, 91058 Erlangen, Germany}

\affiliation{Max Planck Institute for the Science of Light, Günther-Scharowsky-Straße
1 Bau 24, 91058 Erlangen, Germany}

\affiliation{Department of Physics, Technical University of Denmark, Fysikvej,
2800 Kongens Lyngby, Denmark}

\author{F. Marquardt}

\affiliation{Institute for Theoretical Physics, Friedrich-Alexander-Universität
Erlangen-Nürnberg, Staudtstr. 7, 91058 Erlangen, Germany}

\affiliation{Max Planck Institute for the Science of Light, Günther-Scharowsky-Straße
1 Bau 24, 91058 Erlangen, Germany}
\begin{abstract}
It has been predicted and experimentally demonstrated that by injecting
squeezed light into an optomechanical device it is possible to enhance
the precision of a position measurement. Here, we present a fundamentally
different approach where the squeezing is created directly inside
the cavity by a nonlinear medium. Counterintuitively, the enhancement
of the signal to noise ratio works by de-amplifying precisely the
quadrature that is sensitive to the mechanical motion without losing
quantum information. This enhancement works for systems with a weak
optomechanical coupling and/or strong mechanical damping. This could
allow for larger mechanical bandwidth of quantum limited detectors
based on optomechanical devices. Our approach can be straightforwardly
extended to Quantum Non Demolition (QND) qubit detection.
\end{abstract}
\maketitle
Recent progress in cavity optomechanics \cite{Aspelmeyer2014,Metcalfe2014}
has been so exceptional that the precision of a position measurement
has been pushed until the limit set by the principles of quantum mechanics,
the so-called Standard Quantum Limit (SQL) \cite{Caves1980,Clerk2010,BraginskyandKhalilibook}.
A measurement precision close to the SQL has been demonstrated in
optomechanical devices with cavities both in the optical \cite{Anetsberger2009,Purdy15022013,Wilsonpreprint2014}
and in the microwave \cite{Teufel2009} domain. Optomechanical position
detection is not only of fundamental interest but finds also application
in acceleration \cite{Krause2012,Cervantes2014}, magnetic field \cite{Forstner2012,Forstner2012a},
and force detectors \cite{Gavartin2011,Liu2012}. Thus, an important
goal for the future is to develop new techniques to enhance its precision
on different optomechanical platforms. Seminal efforts have focused
on gravitational wave detection in optomechanical interferometers
\cite{Caves1981,Unruh1983,Bondurant1986,Pace1993,Rehbein2005}. The
standard route to enhance the detection precision consists in injecting
squeezed light into the interferometer  \cite{Caves1981,Unruh1983,Collett1984}.
This technique has recently been demonstrated in the Laser Interferometer
Gravitational Wave Observatory (LIGO) \cite{AasiJ.2013} and in a
cavity optomechanics setup \cite{Hoff2013}. Externally generated
squeezed light could also find application in QND qubit state detection
\cite{Barzanjeh2014,Didier2015}. Alternatively, one can enhance a
dispersive quantum measurement by generating the appropriate squeezing
directly inside the cavity by means of a Kerr nonlinearity  \cite{Bondurant1986,Collett1984,Nation2008,Rehbein2005,Laflamme2011,Laflamme2012},
by the dissipative optomechanical interaction \cite{Kronwald2014},
or potentially, by exploiting the ponderomotive squeezing \cite{Brooks2012,Safavi-Naeini2013,Purdy2013}.

\begin{figure}
\includegraphics[width=0.9\columnwidth]{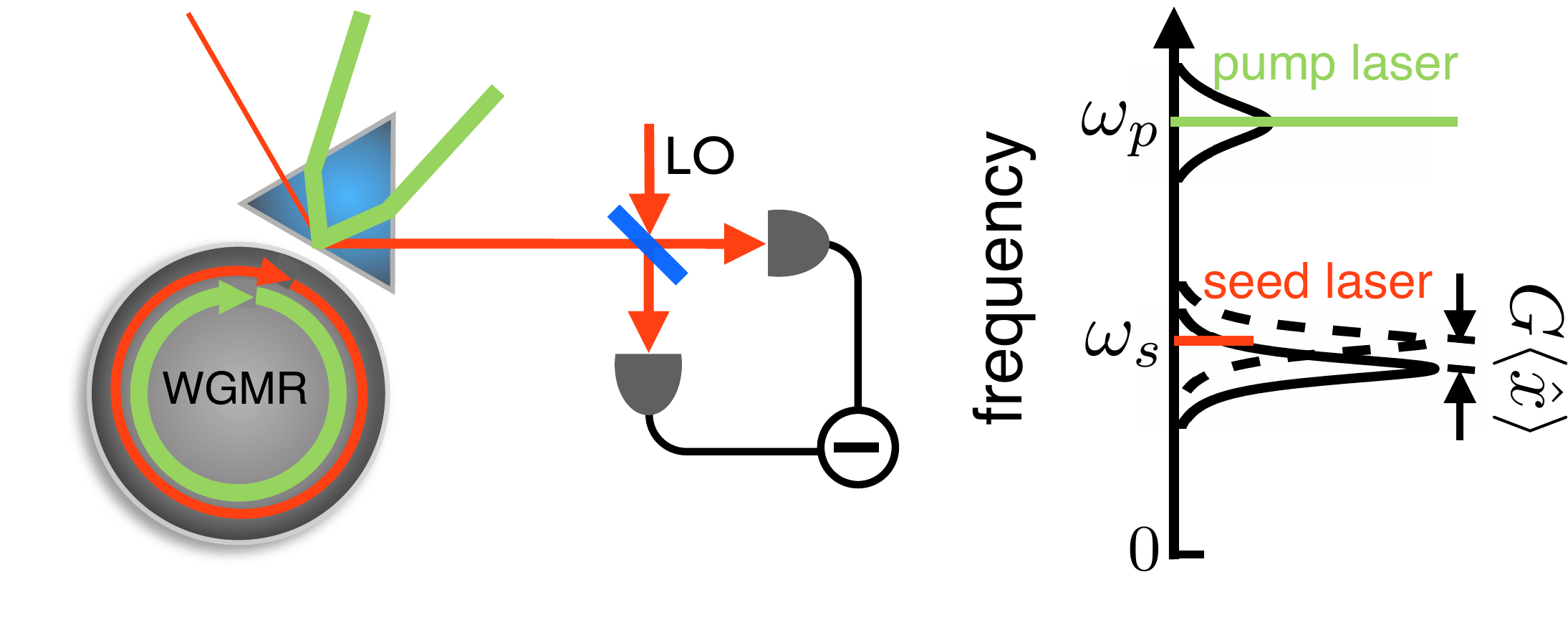}\protect\caption{Setup for parametrically amplified optomechanical position measurement.
A whispering gallery mode resonator (WGMR) with a $\chi^{(2)}$ nonlinearity
has two modes, with eigenfrequencies $\omega_{s}$ (signal) and $\omega_{p}=2\omega_{s}$
(pump). A laser at the pump mode resonance squeezes the fluctuations
of the signal mode. An additional laser beam with the appropriate
phase and frequency $\omega_{s}$ (seed laser) is amplified by the
cavity. The signal mode resonance depends on the amplitude $\langle\hat{x}\rangle$
of a mechanical mode deformation. Thus, the mechanical vibrations
are imprinted in the seed laser phase shift, detected in a homodyne
setup. \label{fig:setup}}
\end{figure}

In this letter, we propose a new pathway to precision enhancement
in optomechanical detection. In our approach, a nonlinear cavity is
operated as a phase-sensitive parametric amplifier, as shown in Fig.~\ref{fig:setup}.
It amplifies a seed laser beam and its intensity fluctuations. Simultaneously,
it \emph{de-amplifies} the phase quadrature where the mechanical vibrations
are imprinted. At first sight it might appear counter-intuitive that
de-amplification can improve a (quantum) measurement. Here, we suggest
that it might be worth to de-amplify a signal if the noise is suppressed
by a larger factor thus obtaining a net enhancement of the signal
to noise ratio. Indeed, our analysis shows that for optomechanical
position detection a de-amplification of the phase quadrature induces
only a limited suppression of the signal but simultaneously can strongly
suppress the measurement noise. Our scheme could be implemented using
a crystalline whispering gallery mode resonator \cite{Savchenkov2004}.
Such devices offer a well established platform for optomechanics \cite{Hofer2010,Ding2010,Aspelmeyer2014,Metcalfe2014}.
Resonators with an optical $\chi^{(2)}$ nonlinearity can be operated
as parametric amplifiers in the quantum regime \cite{Furst2011,Fortsch2013}.
The exciting perspective of an interplay of optical and optomechanical
nonlinearities has already inspired a few theoretical investigations
\cite{Huang2009,XuerebPaternostro2012,Benllochpreprint2014,Noripreprint2014}.
Alternative implementations of our scheme include optomechanical crystals
\cite{Eichenfield2009}, when made out of nonlinear materials such
as AlN \cite{Bochmann2013} and a Josephson parametric amplifier \cite{Castellanos-Beltran2007}
coupled to a mechanical membrane or a qubit.

We consider a degenerate parametric amplifier which is tuned to have
a pair of modes with frequencies $\omega_{p}$ and $\omega_{s}$ ($p$
pump, $s$ signal) where $\omega_{p}=2\omega_{s}$, and the pump mode
is driven resonantly. In the following, we denote as $\kappa_{s}$
and $\kappa_{p}$ the decay rates of the corresponding cavity modes.
We first describe our proposal by considering the standard description
of an ideal degenerate parametric amplifier where the pump mode has
been already adiabatically eliminated and we neglect intrinsic losses.
In a frame rotating at frequency $\omega_{s}$, the degenerate parametric
amplifier is described by the standard linearized Hamiltonian  \cite{WallsMilburn_QuantumOptics}
\[
H_{s}=i\hbar\bar{n}_{p}^{1/2}\nu\left(\hat{a}_{s}^{\dagger}\hat{a}_{s}^{\dagger}-\hat{a}_{s}\hat{a}_{s}\right)/2.
\]
Here, $\hat{a}_{s}$ is the ladder operator for the signal mode, $\nu$
is the single-photon optical nonlinearity and $\bar{n}_{p}$ is the
number of photons circulating in the pump mode. The parametric amplifier
is characterized by its pump parameter $\sigma$,
\begin{equation}
\sigma^{2}=\bar{n}_{p}/\bar{n}_{p}^{({\rm thr)}},\quad\bar{n}_{p}^{({\rm thr)}}=\left(\frac{\kappa_{s}}{2\nu}\right)^{2},\label{eq:npthreshold}
\end{equation}
The signal mode reaches the threshold of self-sustained (optical parametric)
oscillations when the photon number circulating in the pump mode equals
$\bar{n}_{p}^{(thr)}$, corresponding to the pump parameter $\sigma=1$.
Below threshold, the cavity behaves as a phase-sensitive amplifier
as discussed above. 

We want to measure the displacement $\hat{x}$ of a mechanical resonator
with eigenfrequency $\Omega$, effective mass $m$, and decay rate
$\Gamma$, see Fig. \ref{fig:setup}. The mechanical resonance could
be internal to the optical resonator (e.g. a breathing mode) or refer
to the vibrations of an external nano-object coupled evanescently.
A displacement $\hat{x}$ induces a shift $-G\hat{x}$ of the signal
mode frequency, described by a Hamiltonian $\hat{H}_{OM}=-\hbar G\hat{a}_{s}^{\dagger}\hat{a}_{s}\hat{x}$
\cite{Aspelmeyer2014}. We assume here that the effects of the coupling
between the displacement and the pump mode are negligible (otherwise
the additional information contained in the phase shift of the reflected
pump beam would also have to be monitored). We measure the displacement
$\hat{x}$ by extracting the output signal phase of a seed drive injected
at the bare cavity resonance $\omega_{s}$, where it is most sensitive
to the jittering of the optical resonance induced by the mechanical
vibrations. When the seed laser injects a large number $\bar{n}_{s}$
of circulating photons into the signal mode (below, we specify this
condition more precisely), we can linearize the optomechanical interaction
\cite{Aspelmeyer2014}. Then the mechanical vibrations couple to the
optical field quadrature $\hat{X}=(\hat{a}_{s}+\hat{a}_{s}^{\dagger}-2\sqrt{n_{s}})/\sqrt{2}$
describing the amplitude fluctuations: $\hat{H}_{OM}=-\hbar G\sqrt{2\bar{n}_{s}}\hat{X}\hat{x}$.
We arrive at the Langevin equations for the optical signal mode quadratures
$\hat{X}$ (amplitude) and $\hat{Y}$ (phase): 
\begin{eqnarray}
\dot{\hat{X}} & = & -(1-\sigma)\kappa{}_{s}\hat{X}/2+\sqrt{\kappa_{s}}\hat{X}^{(in)}\nonumber \\
\dot{\hat{Y}} & = & -(1+\sigma)\kappa{}_{s}\hat{Y}/2+\sqrt{2\bar{n}_{s}}G\hat{x}+\sqrt{\kappa{}_{s}}\hat{Y}^{(in)}.\label{eq:langevin}
\end{eqnarray}
Here, we have set all absorptive losses to zero (more on that later).
We defined $\hat{Y}=i(\hat{a}_{s}^{\dagger}-\hat{a}_{s})/\sqrt{2}$,
and $\hat{X}^{(in)}$ and $\hat{Y}^{(in)}$ are the standard vacuum
input fields (the quantum fluctuations of the laser beam at the input)\cite{WallsMilburn_QuantumOptics}.
As seen in Eq.~(\ref{eq:langevin}), the presence of the nonlinear
medium and the pump drive manifests itself in the de-amplification
of the phase quadrature and a corresponding amplification of the amplitude
quadrature. In the limit $\sigma\to0$, we recover the Langevin equations
for a cavity measuring the mechanical displacement in the standard
approach without squeezing. 

\begin{figure}
\includegraphics[width=1\columnwidth]{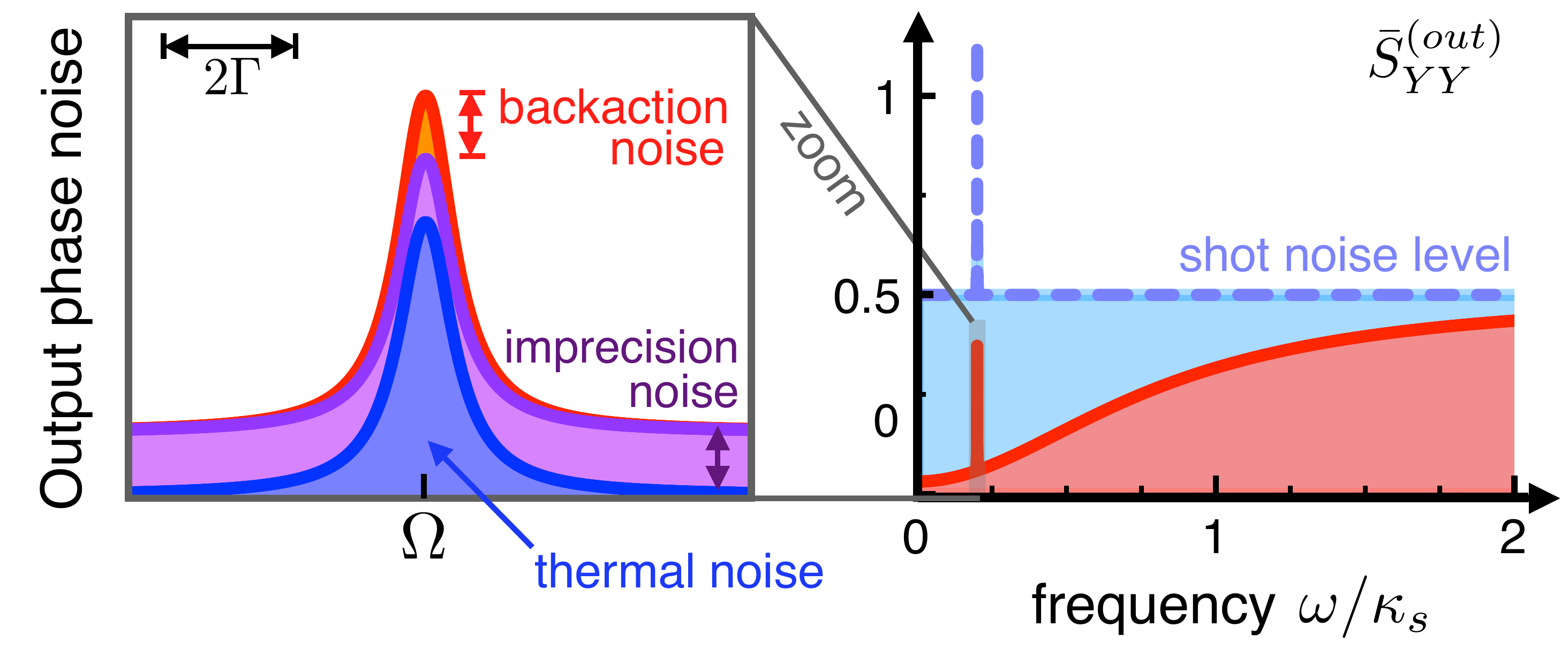}\protect\caption{\label{fig:noisevsfrequency}Output phase noise $\bar{S}_{YY}^{(out)}$
as a function of frequency. Comparison between the phase noise in
presence and in absence of the pump drive for the same number of circulating
photons $\bar{n}_{s}$. In presence of the pump laser (pump parameter
$\sigma=0.6$), the background noise inside the amplifier bandwidth
is squeezed below the shot noise level by more than 3dB. The signal
amplitude is also reduced, but in this case the reduction is bounded
by the 3dB limit. The number of circulating photons $\bar{n}_{s}$
is chosen to yield the minimum added noise allowed by the SQL, for
$\sigma=0.6$. Thus, the imprecision noise and the backaction noise
(shown in the zoom) have the same intensity at the mechanical resonator
eigenfrequency $\Omega$. The remaining parameters are: $\Omega=0.2\kappa_{s}$,
$\Gamma=10^{-3}\kappa_{s},$ $k_{B}T/\hbar\Omega=1$.}
\end{figure}

To improve a measurement by de-amplification might not sound promising.
The measurement noise will be de-amplified but one could reasonably
expect that this effect will be offset by the de-amplification of
the signal. Indeed, it is true that the response of the cavity to
both the vacuum noise and the mechanical vibrations is decreased by
the same factor. From Eq.~(\ref{eq:langevin}), the intracavity phase
quadrature in frequency space is 
\begin{equation}
\hat{Y}[\omega]=\chi_{Y}(\omega)\left(\sqrt{2\bar{n}_{s}}G\hat{x}[\omega]+\sqrt{\kappa{}_{s}}\hat{Y}^{(in)}[\omega]\right)
\end{equation}
 with the intracavity susceptibility
\[
\chi_{Y}=[-i\omega+(1+\sigma)\kappa_{s}/2]^{-1}.
\]
 We note in passing that the largest possible suppression, a factor
of $2$, occurs in the limit $\omega\to0$ and $\sigma\to1$. This
is the well known $3$dB limit of intracavity squeezing \cite{MILBURN1981}.
However, the suppression of the background noise and of the mechanical
signal is different outside the cavity. From the input/output relation
$\hat{Y}^{({\rm out)}}=\hat{Y}^{({\rm in)}}-\sqrt{\kappa_{s}}\hat{Y}$
we find 
\begin{equation}
\hat{Y}^{(out)}[\omega]=[1-\kappa_{s}\chi_{Y}(\omega)]\hat{Y}^{(in)}[\omega]-\sqrt{2\kappa_{s}\bar{n}_{s}}G\chi_{Y}(\omega)\hat{x}[\omega]\label{eq:outputfield}
\end{equation}
From this formula we see that the response of the phase quadrature
of the transmitted signal to the mechanical vibrations is still governed
by the intracavity susceptibility and is thus subject to the $3$dB
limit of squeezing. In contrast, the output phase noise is squeezed
below the $3$dB limit by the destructive interference between the
reflected input noise and the response of the cavity to that noise.
Indeed, it is well-known that the output noise squeezing can be arbitrarily
large \cite{Collett1984}. Thus, we expect an overall enhancement
of the measurement precision accompanied by de-amplification. This
behavior is displayed by the symmetrized spectral density of the output
phase quadrature 
\begin{equation}
\bar{S}_{YY}(\omega)=\int_{-\infty}^{\infty}\frac{dt}{2}e^{i\omega t}\langle\big\{\hat{Y}^{(out)}(t),\hat{Y}^{(out)}(0)\big\}\rangle,
\end{equation}
i. e. the quantity measured in the homodyne set up, see  Fig.~\ref{fig:noisevsfrequency}. 

We briefly comment on the similarities between our scheme and a nonlinear
cavity operated close to its static bistability \cite{Rehbein2005,Nation2008,Laflamme2011}.
Formally, such a cavity is equivalent to an effectively detuned DPA
\cite{Laflamme2011}. Due to the effective detuning, the amplitude
and phase fluctuations become correlated. An important consequence
of such correlations is that the SQL  is reached only away from the
mechanical resonance (whereas for us it is reached precisely at resonance)
\cite{Laflamme2011}. For such a quantum-limited measurement, one
should measure the quadrature whose homodyne signal is amplified by
the cavity \cite{Laflamme2011}. If one does not aim at a quantum-limited
measurement, one can also measure the de-amplified quadrature \cite{Rehbein2005}.
This leads to an improvement of the signal-to-noise ratio similar
to the one observed here. However, this is accompanied by a loss of
quantum efficiency because most of the information regarding the mechanical
vibrations is  imprinted on the other, amplified quadrature, and thus
the SQL would not be reached. Most importantly, there is not such
a trade off in our scheme where all information is imprinted on the
de-amplified quadrature. 
In order to quantify the net enhancement of the measurement precision,
it is convenient to define the measured noise \emph{referred back
to the input $\bar{S}_{xx}^{(meas)}=\bar{S}_{YY}/(2\kappa_{s}\bar{n}_{s}G^{2}|\chi_{Y}|^{2})$}.
Then, from Eq.~(\ref{eq:outputfield})\emph{ }the measured noise
takes the form $\bar{S}_{xx}^{(meas)}(\omega)=\bar{S}_{xx}(\omega)+\bar{S}_{xx}^{(add)}$
where $\bar{S}_{xx}(\omega)$ describes the symmetrized mechanical
noise in absence of optomechanical backaction, whereas $\bar{S}_{xx}^{(add)}$
is the noise added during the measurement. We are interested in the
noise at frequency $\Omega$ where the mechanical spectrum is peaked.
Since there is a typical number of circulating photons (specific of
the device) that can be tolerated without inducing strong heating
effects, we use as a figure of merit of our measurement scheme the
added noise $\bar{S}_{xx}^{(add)}(\Omega)$ for a \emph{fixed total
number of circulating photons,} $\bar{n}=\bar{n}_{s}+\bar{n}_{p}$.
Below, we show that internally generated optical squeezing can strongly
enhance the precision for optomechanical couplings that are small
compared to the optical nonlinearities, when 
\begin{equation}
{\cal C}_{thr}\equiv\frac{g_{0}^{2}\kappa_{s}}{\Gamma\nu^{2}}=\frac{4g_{0}^{2}\bar{n}_{p}^{(thr)}}{\Gamma\kappa_{s}}\ll1.\label{eq:thresholdcooperativity}
\end{equation}
Here, $g_{0}=G/x^{{\rm ZPF}}$ is the single-photon optomechanical
coupling (the optical frequency shift by a single phonon) \cite{Aspelmeyer2014}.
The parameter ${\cal C}_{thr}$, which we refer to as \emph{threshold
cooperativity,} is the optomechanical cooperativity if $\bar{n}_{s}=\bar{n}_{p}^{(thr)}$
photons were in the signal mode. It quantifies the ratio of optomechanical
and nonlinear coupling. Notice that in the absence of squeezing the
SQL is reached for the optomechanical cooperativity ${\cal C}=1/4$
\cite{Aspelmeyer2014}. Thus, if ${\cal C}_{th}\ll1$ it is not possible
to achieve a precision close to the SQL by injecting all available
photons $\bar{n}_{s}\sim\bar{n}_{p}^{(thr)}$ directly into the signal
mode. Instead, one can enhance the measurement precision by injecting
part of the photons into the pump mode to generate squeezing, as shown
below. 

\begin{figure}
\includegraphics[width=1\columnwidth]{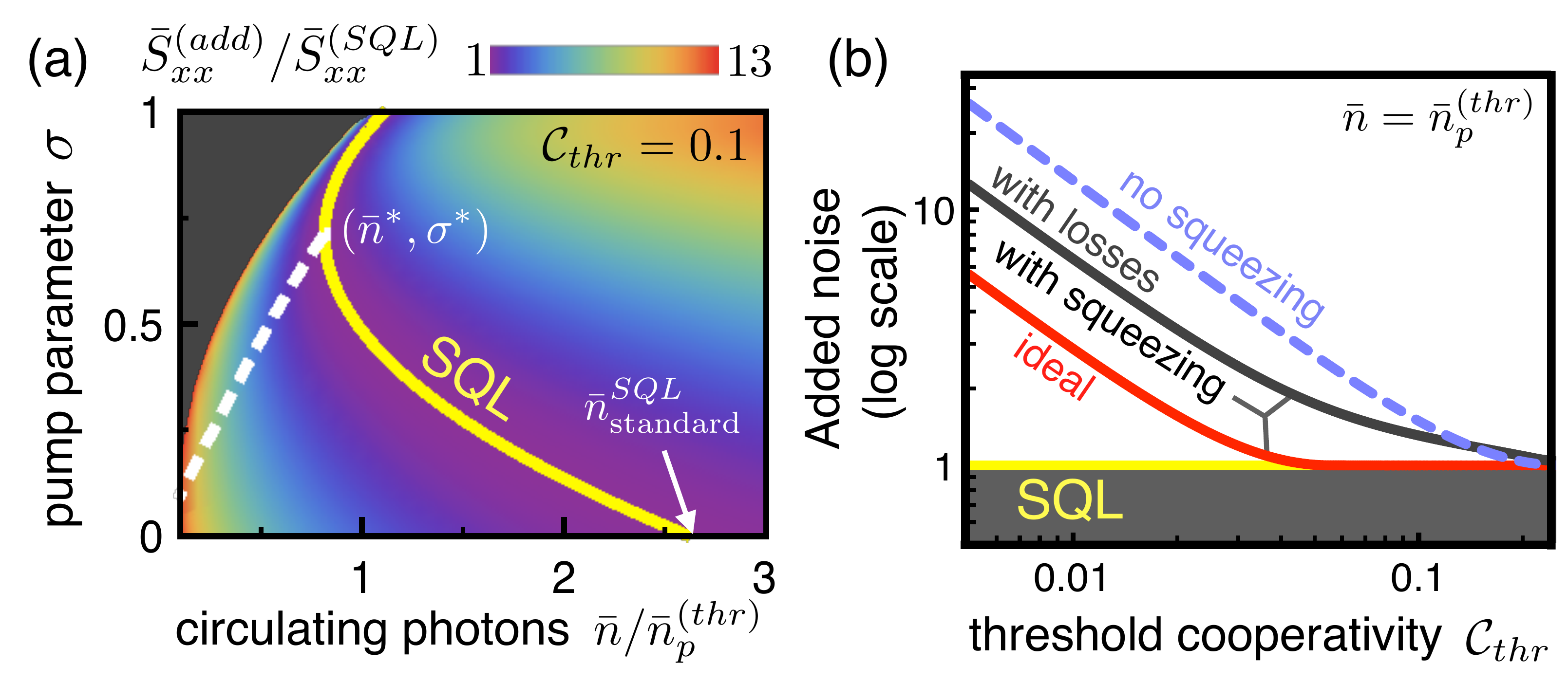}\protect\caption{\label{fig:Added-noise}(a) Added noise at the mechanical frequency
as a function of the total number of circulating photons $\bar{n}=\bar{n}_{s}+\bar{n}_{p}$
and the pump parameter $\sigma$, for ${\cal C}_{th}=0.1$. The coordinates
$(\bar{n}/\bar{n}_{p}^{(thr)},\sigma)$ where the added noise equals
the SQL are indicated by the yellow solid line, see Eq.~(\ref{eq:nSQLofsigma}).
For $\bar{n}<\bar{n}^{*}$ the added noise is always larger than the
SQL. In this case, the minimum noise for a fixed $\bar{n}$ is realized
on the white dashed line, where the pump parameter $\sigma^{(opt)}$
is given by Eq.~(\ref{eq:sigmaopt}). (b) Added noise as a function
of the cooperativity ${\cal C}_{th}$, for $\bar{n}=\bar{n}_{p}^{(thr)}$:
in the ideal case with squeezing ($\kappa^{(abs)}=0,$ $\kappa_{p}/\kappa_{s}\to\infty$
), in the presence of losses and squeezing ($\kappa^{(abs)}=\Omega^{2}/\kappa_{s},$
$\kappa_{p}/\kappa_{s}\to20$ ), and in the absence of squeezing and
losses ($\sigma=0$). For the first two curves, the pump parameter
$\sigma$ is chosen to minimize the added noise. Notice that the damping
due to photon up-conversion $4\nu^{2}\bar{n}_{s}/\kappa_{p}$ depends
on $\sigma$ via $\bar{n}_{s}$, $\bar{n}_{s}=(1-\sigma^{2})\bar{n}_{p}^{(thr)}$.
In both panels we have chosen $\kappa_{s}/\Omega=10$.}
\end{figure}

We now calculate the added noise $\bar{S}_{xx}^{(add)}$. We distinguish
between two different contributions \cite{Caves1980,Clerk2010,BraginskyandKhalilibook}:
the so-called \emph{imprecision noise} $\bar{S}_{xx}^{(imp)}(\omega)$
and \emph{the backaction noise} $\bar{S}_{xx}^{(back)}(\omega)$.
The former is due to the shot noise phase fluctuations. The latter
is the additional mechanical noise induced by the backaction of the
light onto the mechanics. It can be expressed as $\bar{S}_{xx}^{(back)}(\omega)=|\chi_{M}(\omega)|^{2}\bar{S}_{FF}(\omega)$
in terms of the mechanical susceptibility $\chi_{M}(\omega)=m^{-1}(\Omega^{2}-\omega^{2}+i\omega\Gamma)^{-1}$
and the noise spectrum $\bar{S}_{FF}$ of the radiation pressure force
$\hat{F}=\sqrt{2\bar{n}_{s}}\hbar G\hat{X}$. We note in passing that
our measurement scheme could also find application in the detection
of any degree of freedom coupled dispersively to the cavity, e. g.
a qubit \cite{Clerk2010}. From Eq.~(\ref{eq:langevin}) we can readily
derive the identity $\bar{S}_{xx}^{(imp)}(\omega)\bar{S}_{FF}(\omega)=\hbar^{2}/4$
valid for all values of $\sigma$. It is well known that when this
equality holds both the position detection of resonant vibrations
and the QND qubit state detection are quantum limited \cite{Caves1980,Clerk2010,BraginskyandKhalilibook}.
We compute the overall added noise $\bar{S}_{xx}^{(add)}=\bar{S}_{xx}^{(imp)}+\bar{S}_{xx}^{(back)}$
from Eq.~(\ref{eq:langevin}), 
\begin{equation}
\frac{\bar{S}_{xx}^{(imp)}}{\bar{S}_{xx}^{SQL}}=\frac{(1-\sigma)^{2}+4\Omega^{2}/\kappa_{s}^{2}}{8{\cal C}_{thr}(\bar{n}/n_{p}^{(thr)}-\sigma^{2})},\quad\frac{\bar{S}_{xx}^{(back)}}{\bar{S}_{xx}^{SQL}}=\frac{\bar{S}_{xx}^{SQL}}{4\bar{S}_{xx}^{(imp)}}.\label{eq:Simp}
\end{equation}
Here, we have introduced the minimum added noise allowed by the SQL
$\bar{S}_{xx}^{SQL}=\hbar/m\Omega\Gamma$ \cite{Caves1980,Clerk2010,BraginskyandKhalilibook}.
Fig. \ref{fig:Added-noise}(a) shows the added noise Eq.~(\ref{eq:Simp})
as a function of the circulating photon number $\bar{n}$ and the
pump parameter $\sigma.$ For $\sigma=0$ (no circulating photons
in the pump mode), we recover the result for standard optomechanical
detection. The SQL is reached for $\bar{S}_{xx}^{(add)}=2\bar{S}_{xx}^{(imp)}=2\bar{S}_{xx}^{(back)}=\bar{S}_{xx}^{SQL}$
\cite{Caves1980,Clerk2010,BraginskyandKhalilibook}, see also the
zoom in Fig.~\ref{fig:noisevsfrequency}. From Eq.~(\ref{eq:Simp}),
we find the required photon number 
\begin{equation}
\frac{\bar{n}^{SQL}(\sigma)}{\bar{n}_{p}^{(thr)}}=\frac{(1-\sigma)^{2}+4\Omega^{2}/\kappa_{s}^{2}}{4{\cal C}_{thr}}+\sigma^{2}.\label{eq:nSQLofsigma}
\end{equation}
It is shown as a yellow solid line in Fig.~\ref{fig:Added-noise}(a).
By minimizing $\bar{n}^{SQL}(\sigma)$ as a function of $\sigma$,
we find the minimal number of circulating photons $\bar{n}^{*}$ necessary
to reach the SQL and the corresponding optimal pump parameter $\sigma^{*}$,
\begin{equation}
\bar{n}^{*}=\bar{n}^{SQL}(\sigma^{*}),\quad\sigma^{*}=\left(1+4{\cal C}_{thr}\right)^{-1}.\label{eq:nSQLmin}
\end{equation}
Compared to the standard scheme, where the SQL is reached for $\bar{n}_{standard}^{SQL}=\bar{n}^{SQL}(\sigma=0)$
circulating photons, the required number of photons is suppressed
by a factor of 
\begin{equation}
\bar{n}_{standard}^{SQL}/\bar{n}^{*}=\frac{1+4\Omega^{2}/\kappa_{s}^{2}}{1-(4{\cal C}_{thr}+1)^{-1}+4\Omega^{2}/\kappa_{s}^{2}}.
\end{equation}
The suppression factor increases monotonically with increasing optical
nonlinearity (decreasing threshold cooperativity ${\cal C}_{thr}$)
and reaches the asymptotic value $\kappa_{s}^{2}/4\Omega^{2}$ for
large optical nonlinearities (${\cal C}_{thr}\ll1$), in the bad cavity
limit $\Omega\ll\kappa_{s}$. Our method is still useful even when
it is not possible to reach the maximum precision allowed by the SQL
because the typical number of circulating photons tolerated in the
the device is too small (smaller than $\bar{n}^{*}$). In this case,
the added noise remains larger than $\bar{S}_{xx}^{SQL}$, yet it
can still be decreased by the squeezing. By minimizing $\bar{S}_{xx}^{(add)}$
in Eq. (\ref{eq:Simp}) as a function of $\sigma$ for a fixed $\bar{n}$
(smaller than $\bar{n}^{*}$), we find the optimal pump parameter
\begin{equation}
\sigma^{(opt)}=\frac{{\cal B}}{2}-\left(\frac{{\cal B}^{2}}{4}-\frac{\bar{n}}{\bar{n}_{p}^{(thr)}}\right)^{1/2},\quad{\cal B}=1+\frac{\bar{n}}{\bar{n}_{p}^{(thr)}}+4\frac{\Omega^{2}}{\kappa_{s}^{2}}.\label{eq:sigmaopt}
\end{equation}
It increases monotonically with the number of circulating photons
and reaches the value $\sigma=\sigma^{*}$ for $\bar{n}=\bar{n}^{*}$,
see the white dashed line in Fig.~\ref{fig:Added-noise}(a). 

Next we go beyond the ideal description of a parametric amplifier
by considering the effects of losses. Those are potentially deleterious
as they decrease the amount of achievable squeezing. In a parametric
amplifier there are two main loss channels: (i) photon absorption
and (ii) photon up-conversion, where photons which are up-converted
by the $\chi^{(2)}$ interaction $i\hbar\nu(\hat{a}_{s}^{\dagger2}\hat{a}_{p}-h.c.)/2$
decay via the pump mode. This process is enhanced in the presence
of a large number $\bar{n}_{s}$ of signal mode photons. Thus, the
overall loss rate takes the form (see also Appendix \ref{sec:AppendixB}
for a full derivation) 
\begin{equation}
\kappa_{s}^{(loss)}=\kappa_{s}^{(abs)}+4\nu^{2}\bar{n}_{s}/\kappa_{p}.\label{eq:lossrate-1}
\end{equation}
In order to suppress the losses via the pump mode it is therefore
important to have a large pump decay rate $\kappa_{p}$. The effect
of losses is investigated in Fig.~\ref{fig:Added-noise}(b). It shows
that the measurement precision can still be noticeably enhanced by
the squeezing in a broad range of threshold cooperativities ${\cal C}_{thr}$.
Indeed, an explicit calculation shows that the added noise is dramatically
increased only for $\kappa_{s}^{(abs)}\gtrsim\Omega^{2}/\kappa_{s}$,
see Appendix \ref{sec:AppendixB}. A large pump decay rate is helpful
to suppress losses. For a small pump decay rate $\kappa_{p}$, one
could suppress the upconversion losses by introducing a detuning between
the pump laser and the pump mode, while keeping the (pump) seed laser
at (twice) the effective signal mode eigenfrequency. Alternatively,
one could improve the measurement precision by monitoring also the
light scattered by the pump mode. 

The inevitable enhancement of the intensity fluctuations in the proposed
measurement scheme represents a potential contradiction of the assumption
of small fluctuations, inherent to the linearized Langevin equations
(\ref{eq:langevin}). However, it can be shown that the enhanced fluctuations
remain compatible with the linearization, provided that the single-photon
nonlinearity $\nu$ is not too large, $\nu\ll\Omega$, see Appendix
\ref{sec:AppendixA}.

The regime of small threshold cooperativities ${\cal C}_{thr}\ll1$
is realized in state-of-the art lithium-niobate microdisks \cite{Furst2010,Furst2011,Fortsch2013}.
These devices have breathing modes with eigenfrequencies $\Omega$
in the MHz range. Typical single-photon optomechanical couplings are
in the sub-Hz range, whereas single-photon optical nonlinearities
$\nu$ are in the kHz range. Thus, the regime ${\cal C}_{thr}\ll1$
is compatible with the bad cavity limit even for disks with large
mechanical quality factors. Moreover, the nonlinear corrections to
the Langevin equations (\ref{eq:langevin}) will be small.

In conclusion, we have shown that the precision of optomechanical
position detection can be strongly enhanced by de-amplification without
loss of quantum efficency. Our method could pave the way to the quantum
limited position detection of mechanical resonators with larger decay
rates. This would allow faster detection of forces yielding an increase
of the bandwidth of quantum limited detectors based on optomechanical
devices \cite{Metcalfe2014}. A natural extension of our scheme could
find application in QND qubit state detection. 

V. P. and F. M. acknowledge support by an ERC Starting Grant OPTOMECH,
by the DARPA project ORCHID, and by the European Marie-Curie ITN network
cQOM; Ch. M. acknowledges support from the Alexander von Humboldt
Foundation. We thank Aash Clerk, Michele Collodo, Michael Förtsch,
and Andreas Kronwald for discussions.

\appendix

\section{Derivation of the linearized Hamiltonian for the optomechanical degenerate
parametric amplifier\label{sec:AppendixA}}

We start from the standard input-output formalism \cite{WallsMilburn_QuantumOptics}
for a degenerated parametric amplifier (DPMA) coupled via radiation
pressure to an underdamped mechanical oscillator. Both optical modes
are driven resonantly. In a frame where they are rotating at their
respective eigenfrequencies, $\omega_{p}$ and $\omega_{s}=\omega_{p}/2$
the Langevin equation reads
\begin{eqnarray}
\dot{\hat{a}}_{p} & = & -\kappa{}_{p}\hat{a}_{p}/2-\nu\hat{a}_{s}^{2}/2+\sqrt{\kappa_{p}^{(in)}}\hat{a}_{p}^{(in)}+\sqrt{\kappa_{p}^{(abs)}}\hat{a}_{p}^{(abs)}\nonumber \\
\dot{\hat{a}}_{s} & = & ig_{0}\left(\hat{b}+\hat{b}^{\dagger}\right)\hat{a}_{s}-\kappa{}_{s}\hat{a}_{s}/2+\nu\hat{a}_{s}^{\dagger}\hat{a}_{p}+\sqrt{\kappa_{s}^{(in)}}\hat{a}_{s}^{(in)}\nonumber \\
 &  & +\sqrt{\kappa_{s}^{(abd)}}\hat{a}_{s}^{(abs)}\nonumber \\
\dot{\hat{b}} & = & (-i\Omega-\Gamma/2)\hat{b}+ig_{0}\hat{a}_{s}^{\dagger}\hat{a}_{s}+\sqrt{\Gamma}\hat{b}^{(in)}\label{eq:Langevinnonlin}
\end{eqnarray}
Moreover, we denote by $\hat{b}$ and $\hat{a}_{p/s}$ the phonon
and photon annihilation operators, respectively; $\kappa_{p/s}^{(in)}$
and $\kappa_{p/s}^{(abs)}$ are the photon outcoupling and absorption
rates, respectively; Their sums give the overall decay rates $\kappa_{p/s}=\kappa_{p/s}^{(in)}+\kappa_{p/s}^{(abs)}$
; The pump and seed laser have amplitudes $\langle\hat{a}_{p/s}^{(in)}\rangle=\alpha_{p/s}^{(in)}$. 

We use the standard procedure \cite{WallsMilburn_QuantumOptics} to
linearize the Langevin equations (\ref{eq:Langevinnonlin}): we divide
the ladder operators $\hat{b}$ and $\hat{a}_{p/s}$ into the sum
of their average stationary amplitudes and the corresponding fluctuating
fields, $\hat{b}=\langle\hat{b}\rangle+\delta\hat{b}\equiv\beta+\delta\hat{b}$,
$\hat{a}_{p/s}=\langle\hat{a}_{p/s}\rangle+\delta\hat{a}_{p/s}\equiv\alpha_{p/s}+\delta\hat{a}_{p/s}$;
By plugging this decomposition into the full nonlinear Langevin equation
(\ref{eq:Langevinnonlin}) and neglecting all the correlations between
the fluctuating fields $\delta\hat{b}$ and $\delta\hat{a}_{p/s}$,
we arrive to two set of equations for the stationary average fields
and the noise operators. The \emph{classical }equations for the stationary
amplitudes,
\begin{eqnarray*}
-(\kappa{}_{p}\alpha_{p}+\nu\alpha_{s}^{2})/2+\sqrt{\kappa_{p}^{(in)}}\alpha_{p}^{(in)} & = & 0,\\
i2g_{0}^{2}\bar{n}_{s}\alpha_{s}/\Omega-(\kappa{}_{s}\alpha_{s}-2\nu\alpha_{s}^{*}\alpha_{p})/2+\sqrt{\kappa_{s}^{(in)}}\alpha_{s}^{(in)} & = & 0,\\
-i\Omega\beta+ig_{0}\bar{n}_{s} & = & 0,
\end{eqnarray*}
yield the laser light amplitudes $\alpha_{p}^{(in)}$ and $\alpha_{p}^{(in)}$
required to generate a specific combination of intracavity average
light amplitudes $\alpha_{s}$ and $\alpha_{p}$. In our investigation,
we focus on the parameter regime where the precision of the optomechanical
position detection could be enhanced, $\alpha_{p}=\sigma\kappa_{s}/2\nu$
with $0\leq\sigma<1$ and $\alpha_{s}=\bar{n}_{s}^{1/2}$ (thus $\alpha_{s}$
and $\alpha_{p}$ are positive and real). The noise operators dynamics
is governed by the linearized Langevin equations,
\begin{eqnarray*}
\delta\dot{\hat{a}}_{p} & = & -\frac{\kappa{}_{p}}{2}\delta\hat{a}_{p}-\nu\sqrt{\bar{n}_{s}}\delta\hat{a}_{s}+\sqrt{\kappa_{p}^{(in)}}\delta\hat{a}_{p}^{(in)}\\
 &  & +\sqrt{\kappa_{p}^{(abs)}}\hat{a}_{p}^{(abs)},\\
\delta\dot{\hat{a}}_{s} & = & ig_{0}\sqrt{\bar{n}_{s}}(\hat{b}+\hat{b}^{\dagger})+i2g_{0}^{2}\bar{n}_{s}\delta\hat{a}_{s}/\Omega-\frac{\kappa{}_{s}}{2}\delta\hat{a}_{s}\\
 &  & +\sigma\frac{\kappa{}_{s}}{2}\delta\hat{a}_{s}^{\dagger}+\sqrt{\kappa_{s}^{(in)}}\delta\hat{a}_{s}^{(in)}+\sqrt{\kappa_{s}^{(abs)}}\hat{a}_{s}^{(abs)}\\
\delta\dot{\hat{b}} & = & \left(-i\Omega-\frac{\Gamma}{2}\right)\delta\hat{b}+ig_{0}\sqrt{\bar{n}_{s}}\left(\delta\hat{a}_{s}^{\dagger}+\delta\hat{a}_{s}\right)+\sqrt{\Gamma}\hat{b}^{(in)}.
\end{eqnarray*}
 The above equations describe the linearized dynamics beyond the
ideal description of Eq. (\ref{eq:langevin}). Compared to the ideal
description, they include also intrinsic losses, the coupling between
the signal and pump mode fluctuations, and a radiation pressure induced
effective detuning of the signal mode (by $2g_{0}^{2}\bar{n}_{s}/\Omega$). 

Before adding more technical details to the discussion of the main
text regarding the effects of losses and of the pump-signal coupling
we have to explain why we have omitted the effective detuning of the
signal mode in the main text. The effective detuning is negligible
when it is much smaller than the cavity bandwidth $\kappa_{s}$, $\Omega\kappa_{s}\gg g_{0}^{2}\bar{n}_{s}$.
Keeping in mind that the Standard Quantum Limit (SQL) in a linear
cavity is reached for $\Gamma\kappa_{s}=16g_{0}^{2}\bar{n}_{s}$ \cite{Aspelmeyer2014},
we can conclude that the shift is negligible if the circulating photon
number is optimized to reach the best possible precision and, at the
same time, the mechanical quality factor is large (this is typically
the case in WGMRs). Even in the case of mechanical oscillators with
not too large quality factors, it could still be possible to eliminate
the radiation pressure induced detuning by tuning appropriately the
cavity spectrum and the laser frequencies. 

In the absence of detuning, it is most convenient to write the Langevin
equations in terms of quadratures,
\begin{eqnarray}
\dot{\hat{Y}}_{p} & = & -\frac{\kappa{}_{p}}{2}\hat{Y}_{p}-\nu\sqrt{\bar{n}}_{s}\hat{Y}+\sqrt{\kappa_{p}^{(in)}}\hat{Y}_{p}^{(in)}+\sqrt{\kappa_{p}^{(abs)}}\hat{Y}_{p}^{(abs)}\nonumber \\
\dot{\hat{X}}_{p} & = & -\frac{\kappa{}_{p}}{2}\hat{X}_{p}-\nu\sqrt{\bar{n}}_{s}\hat{X}+\sqrt{\kappa_{p}^{(in)}}\hat{X}_{p}^{(in)}+\sqrt{\kappa_{p}^{(abs)}}\hat{X}_{p}^{(abs)}\nonumber \\
\dot{\hat{Y}} & = & -(1+\sigma)\kappa{}_{s}\hat{Y}/2+\nu\sqrt{\bar{n}}_{s}\hat{Y}_{p}+\sqrt{2\bar{n}_{s}}Gx\nonumber \\
 &  & +\sqrt{\kappa_{s}^{(in)}}\hat{Y}^{(in)}+\sqrt{\kappa_{s}^{(abs)}}\hat{Y}^{(abs)}\nonumber \\
\dot{\hat{X}} & = & -(1-\sigma)\frac{\kappa{}_{s}}{2}\hat{X}+\nu\sqrt{\bar{n}}_{s}\hat{X}_{p}+\sqrt{\kappa_{s}^{(in)}}\hat{X}^{(in)}\nonumber \\
 &  & +\sqrt{\kappa_{s}^{(abs)}}\hat{X}^{(abs)}\nonumber \\
\dot{\hat{x}} & = & \frac{\hat{p}}{m}-\frac{\Gamma}{2}\hat{x}+\sqrt{\Gamma}\hat{x}^{(in)}\nonumber \\
\dot{\hat{p}} & = & -m\Omega^{2}\hat{x}-\frac{\Gamma}{2}\hat{p}+\sqrt{2\bar{n}_{s}}\hbar G\hat{X}+\sqrt{\Gamma}\hat{p}^{(in)}\label{eq:Langevinnodetuning}
\end{eqnarray}
 Here, $\hat{X}=(\delta\hat{a}_{s}+\delta\hat{a}_{s}^{\dagger})/\sqrt{2}$
and $\hat{Y}=-i(\delta\hat{a}_{s}+\delta\hat{a}_{s}^{\dagger})/\sqrt{2}$
describe the signal mode intensity and phase fluctuations. Analogous
definitions apply to the pump quadratures $\hat{X}_{p}$ and $\hat{Y}_{p}$
and the noise operators $\hat{X}^{(in)},$ $\hat{X}^{(abs)}$, $\hat{Y}^{(in)},$
$\hat{Y}^{(abs)}$ $\hat{X}_{p}^{(in)},$ $\hat{X}_{p}^{(abs)}$,
$\hat{Y}_{p}^{(in)},$ and $\hat{Y_{p}}^{(abs)}\!\!\!$ . Moreover,
$\hat{p}=(m\Omega\hbar/2)^{1/2}(\delta b+\delta b^{\dagger})$ is
the oscillator momentum and $\hat{x}=(\hbar/2m\Omega)^{1/2}(\delta b+\delta b^{\dagger})$
is the displacement counted off from the stationary position, $\hbar G\bar{n}_{s}/m\Omega^{2}$.
We note in passing that the form of the mechanical damping, yielding
a dissipation of the mechanical energy which is independent from the
phase of the vibrations, is consistent with the assumption of underdamped
vibrations decaying over many cycles, $\Gamma\ll\Omega$.

\subsection{Stability analysis}

We have verified that the classical solution $\alpha_{p}=\sigma\kappa_{s}/(2\nu)$
and $\alpha_{s}=\sqrt{\bar{n}_{s}}$ with $\sigma<1$ around which
we are linearizing the dynamics is an attractor. We have computed
the complex eigenvalues $-i\omega_{j}$ ($j=1,6$) of the homogeneous
part of Eq. (\ref{eq:Langevinnodetuning}). The imaginary part $-2{\rm Im}\omega_{j}$
is the intensity decay rate of the corresponding excitations. The
solution is stable if all decay rates are positive. The analytical
solution shows that that the classical solution for the full optomechanical
DPMA is always stable in the parameter regime where the ideal DPMA
is stable, $\alpha_{p}=\sigma\kappa_{s}/(2\nu)$ and $\alpha_{s}=\sqrt{\bar{n}_{s}}$
with $|\sigma|<1$. This result has a simple explanation: In the absence
of optical detuning, there is no optically induced mechanical (anti)damping
and the decay rate of the phonons is always $\Gamma$. Instead, the
coupling between the pump and the signal fluctuations creates an admixture
between these two optical modes and influences the optical decay rates.
However, the smallest decay rate is increased by the beam-splitting
type interaction. Consequently, the instability threshold is shifted
toward larger values of the pump parameter, $|\sigma^{(thr)}|>1$.

\subsection{Analysis of the limit of validity of the nonlinear corrections}

We have also verified that the linearization of the Langevin equations
is a good approximation for the parameter regime compatible with state
of the art devices. This is true even for small threshold cooperativities
${\cal C}_{thr}\ll1$ where the amplifier is operated close to threshold
 ($\sigma\approx1$) and the intensity fluctuations are strongly enhanced,
$\langle\hat{X}^{2}\rangle=(1-\sigma)^{-1}/2$. We consider the case
where the minimum added noise allowed by the SQL is realized in presence
of the smallest possible number of circulating photons $\bar{n}^{*}$
for the optimal pump parameter $\sigma^{*}$, see Eqs. (7,8) and Fig.
3(a) of the main text. We require that the average number of additional
photons due to the fluctuations $\langle\hat{X}^{2}+\hat{Y}^{2}\rangle/2$
is comparatively small, $\langle\hat{X}^{2}+\hat{Y}^{2}\rangle/2\ll\bar{n}_{s}$.
From Eqs. (7,8) we find that in the limit ${\cal C}_{thr}\ll(\Omega/\kappa_{s})^{2}$,
where this constraint is more stringent, it amounts to an upper bound
for the optical nonlinearity, $\nu\ll\Omega$. Thus, the linearized
Langevin equations accurately describe state of the art lithium-niobate
microdisks \cite{Furst2010} where typical optical nonlinearities
are in the kHz range and typical frequencies of vibrational breathing
modes are in the MHz range.

\section{Technical details regarding the calculation of the added noise in
presence of losses\label{sec:AppendixB}}

In Fig. 3(b), we have computed directly from the system of equations
(\ref{eq:Langevinnodetuning}) the added noise referred back to the
input and minimized over the pump parameter $\sigma$. We found that
a rather large pump parameter was necessary to keep the added noise
small. In the regime of large pump decay rates, $\kappa_{p}\gg\kappa_{s},\nu\sqrt{\bar{n}_{s}}$,
it is possible to eliminate adiabatically the pump mode and we arrive
to the Langevin equations
\begin{eqnarray*}
\dot{\hat{Y}} & = & -(1+\sigma)\kappa{}_{s}\hat{Y}/2+\sqrt{2\bar{n}_{s}}G\hat{x}+\sqrt{\kappa_{s}^{(in)}}\hat{Y}^{(in)}\\
 &  & +\sqrt{\kappa_{s}^{(loss)}}\hat{Y}^{(loss)},\\
\dot{\hat{X}} & = & -(1-\sigma)\kappa{}_{s}\hat{X}/2+\sqrt{\kappa_{s}^{(in)}}\hat{X}^{(in)}+\sqrt{\kappa_{s}^{(loss)}}\hat{X}^{(loss)},\\
\dot{\hat{x}} & = & \hat{p}/m-\Gamma/2\hat{x}+\sqrt{\Gamma}\hat{x}^{(in)},\\
\dot{\hat{p}} & = & -m\Omega^{2}\hat{x}-\Gamma/2\hat{p}+\sqrt{2\bar{n}_{s}}\hbar G\hat{X}+\sqrt{\Gamma}\hat{p}^{(in)}.
\end{eqnarray*}
In this regime, the coupling to the pump fluctuations merely increases
the signal mode losses, $\kappa^{(loss)}=\kappa^{(abs)}+4\nu^{2}\bar{n}_{s}/\kappa_{p}$
($\kappa_{s}=\kappa^{(in)}+\kappa^{(loss)}$). Physically any up-converted
photon leaks out of the cavity before it can be down-converted again.
In the adiabatic approximation, the imprecision and backaction noise
become
\begin{eqnarray}
\frac{\bar{S}_{xx}^{(imp)}}{\bar{S}_{xx}^{SQL}} & = & \frac{[(1-\sigma)\kappa_{s}-\kappa_{s}^{(loss)}]^{2}+4[\kappa_{s}^{(in)}\kappa_{s}^{(loss)}+\Omega^{2}]}{8{\cal C}_{th}(\bar{n}/\bar{n}_{p}^{(thr)}-\sigma^{2})\kappa_{s}^{2}},\nonumber \\
\frac{\bar{S}_{xx}^{(back)}}{\bar{S}_{xx}^{SQL}} & = & \frac{2{\cal C}_{thr}(\bar{n}/\bar{n}_{p}^{(thr)}-\sigma^{2})}{(1-\sigma)^{2}+4(\Omega/\kappa_{s})^{2}}.\label{eq:Simp-1}
\end{eqnarray}
For the parameters of Fig. \ref{fig:Added-noise}b, the minimal added
noise calculated from these expressions fits well (it can not be distinguished
with the bare eye) with the result obtained directly from Eqs.~(\ref{eq:Langevinnodetuning}).
From the simple analytical formula in Eq. (\ref{eq:Simp-1}), we see
that the imprecision noise is very sensitive to the losses. Even in
the case where the signal mode is overcoupled $\kappa_{s}^{(in)}\gg\kappa_{s}^{(loss)}$,
the increase can be substantial if $\kappa_{s}\kappa_{s}^{(loss)}\gg\Omega^{2}$.
We roughly estimate the loss rate $\kappa_{s}^{(loss)}$ that can
be tolerated without strongly affecting the measurement precision
in our scheme to be $\kappa_{s}^{(loss)}\lesssim\Omega^{2}/\kappa_{s}.$

\end{document}